\documentclass[journal,comsoc,twocolumn]{IEEEtran}

\ifCLASSINFOpdf
  
\else
  
\fi

\usepackage{ifpdf}
\usepackage{cite}
\usepackage{amsmath}
\usepackage{algorithmic}
\usepackage{algorithm}
\usepackage{multicol}
\usepackage{array}

\usepackage{amssymb,amsfonts}
\usepackage{graphicx}
\usepackage{textcomp}
\usepackage{xcolor}
\usepackage{subcaption}
\usepackage[dvips]{epsfig}
\usepackage{kotex}
\begin{document}

\title{Flight Sensor Data and Beamforming based Integrated UAV Tracking with Channel Estimation using Gaussian Process Regression}

\author{ Ha-Lim Song and Young-Chai Ko
\thanks{
$^{\star}$ H.-L. Song and Y.-C. Ko are with Korea University, Seoul, Korea. Y.-C. Ko is a corresponding author (Email: \{hhhalims,koyc\}@korea.ac.kr).
}}

\maketitle

\begin{abstract}
With explosively increasing demands for unmanned aerial vehicle (UAV) applications, reliable link acquisition for serving UAVs is required. Considering the dynamic characteristics of UAV, it is hugely challenging to persist a reliable link without beam misalignment. 
In this paper, we propose a flight sensor data and beamforming signal based integrated UAV tracking scheme to deal with this problem. The proposed scheme provides a compatible integrated system considering the practical specification of the flight sensor data and the beamforming pilot signal. The UAV position tracking is comprised of two steps: 1) UAV position prediction by the flight sensor data and 2) position update with the beamforming signal using Gaussian process regression (GPR) method, which is a nonparametric machine learning.
The flight sensor data can assist ground station (GS) or UAV nodes in designing the precoding and the receive beamforming matrix with drastically reduced overheads.
The beamforming signal can accomplish high beamforming gain to be maintained even when the flight sensor data is absent. Therefore, the proposed scheme can support the moving target continuously by utilizing these two signals. The simulation results are provided to confirm that the proposed scheme outperforms other conventional beam tracking schemes.
We also derive 3-dimensional (3D) beamforming gain and spectral efficiency (SE) from the mean absolute error (MAE) of the angular value estimation, which can be used as beamforming performance metrics of the data transmission link in advance.
\end{abstract}


\IEEEpeerreviewmaketitle

\section{Introduction}
Unmanned aerial vehicles (UAVs) have a plethora of applications such as
surveillance, rescue, geological exploration, and delivery in civilian and military fields \cite{zeng2016wireless}. Moreover, UAV-enabled communications are also drawing attention as a significant technology in beyond 5G (or 6G) cellulalr systems because UAVs of which location is flexible, are suitable to provide the high data rate and the low latency transmission \cite{zeng2018cellular,zhang2019iot}. For example, UAVs can be used as mobile base stations, relays, and backhaul links in air-to-ground (A2G) and aeiral networks \cite{hayat2016survey,liu2020opportunistic}. In particular, a hierarchical aerial network consists of multi-rotors, fixed-wing vehicles, high-altitude platforms (HAPs), low-orbit (LEO) satellites, and geosynchronous equatorial orbit (GEO) satellites \cite{xiao2016enabling,zhao2018beam}, which is regarded as a promising next-generation network. Various flying vehicles can expand the coverage of communication by establishing ubiquitous connections from ground nodes to satellites \cite{huang2019survey}. To fulfill those tasks, it is essential to acquire a reliable link between the node (or UAV) and UAV. For reliable link with UAV, it is necessary to have an accurate channel and position estimations which can be used for designing the 3-dimensional (3D) beamforming.

On the other hand, the conventional cellular and WLAN systems estimate the channel information from beam management operation \cite{3gpp.38.211,ieee2007ieee}. For example, in 3GPP 5G cellular standard, the beam sweeping scheme based on the discrete Fourier transform (DFT) matrix is adopted for the beam training \cite{giordani2018tutorial}. Similarly, in IEEE802.11ad/ay standard, the beam training consists of two steps such as the coarse beam sweeping and beam refinement \cite{lee2011low}.
Based on those conventional beam managements, the beam tracking schemes are proposed in \cite{deng2019channel,yang2019mobile,lin2019sensor,larew2019adaptive,zhang2018beam,zhang2020uav,zhong2020novel,huang20203d}. 
In \cite{deng2019channel,yang2019mobile,lin2019sensor,larew2019adaptive}, the Bayesian and Kalman filtering based estimation technique is presented for the beam tracking, while in \cite{zhang2018beam}, IEEE 802.15 standard based beam training scheme is proposed to maximize the bandwidth efficiency for UAV. Moreover, in \cite{zhang2020uav,zhong2020novel,huang20203d}, codebook beamforming based scheme is investigated for UAV beam training. Note that all the aforementioned beam training (or tracking) schemes are based on information from communications, not using any other sensor data which is available at UAV.

\begin{figure*}[htbp]
    \centering
    \includegraphics[width=0.8\textwidth]{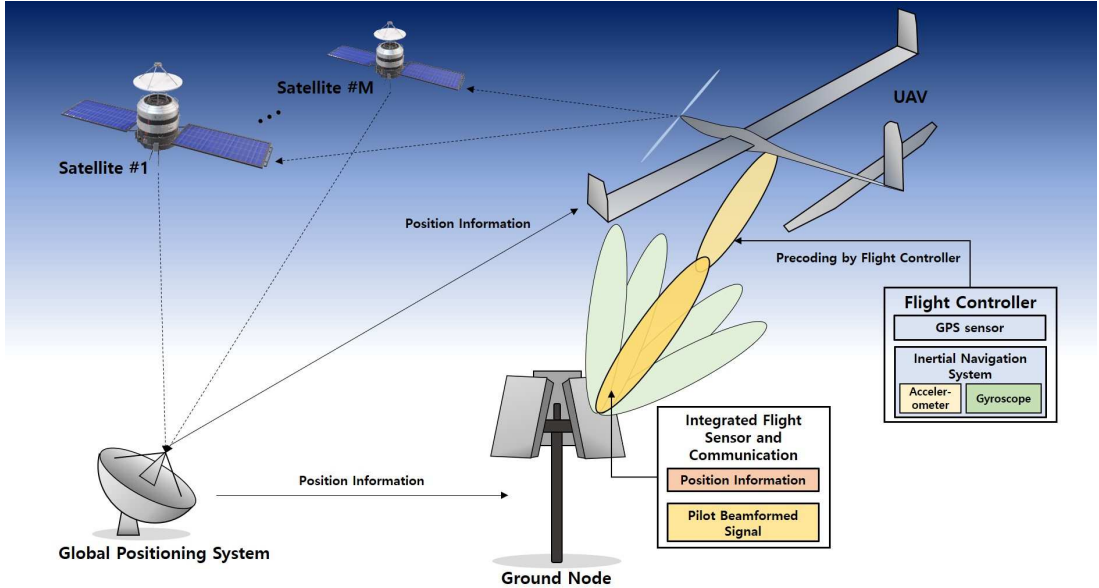}
    \caption{Flight sensor data and beamforming based UAV tracking scenario model.}
    \label{fig:scenario}
\end{figure*}

In flight control framework, control system tracks the target with radars and flight sensors such as global positioning system (GPS), inertial navigation system (INS), and an on-board camera \cite{nemra2010robust,tisdale2008multiple}. INS refers to inertial measurement unit (IMU) sensor with calibration processing. Furthermore, UAV is in general equipped with embedded GPS/INS (EGI). To support the moving UAV with directional beamforming, the spatial beam direction of signal from UAV is required, which can be derived from the position of UAV. Thus, the GPS signal can reduce the beam training overhead drastically. In addition, the spatial beam angle of on-board antenna array varies depending on the attitude status from the perturbation and navigation of UAV, which are main factors for beam misalignment. Therefore, we need to exploit GPS and INS information to reduce the beamforming overhead and beam pointing errors.
There are some works using the sensor data for the UAV communications.

In sensor aided UAV communication researches, there are some achievements of combining the sensor information and the beamforming signal. In \cite{zhao2020doa,miao2019position,miao2020lightweight}, the authors have focused on multiple signal classification (MUSIC) based tracking scheme to obtain the optimal beamforming weight using the GPS signals. However, the MUSIC algorithm with high computational complexity for beam tracking is not appropriate to the analog (or hybrid) beamforming systems, which are more attractive for UAV due to the power consumption and low-cost issues \cite{basha2013beam,zeng2019accessing}.
Moreover, as an alternative, the codebook based beam tracking schemes are discussed in \cite{zhang2020codebook}. The author of \cite{zhang2020codebook} provides a new conformal array based tracking scheme with flight sensors. Furthermore, the control based tracking schemes with flight sensors are presented in \cite{zhao2018beam,zhao2018channel,fu2018multi}. The author of \cite{zhao2018beam} investigates the mechanical beam control correction algorithm using a micro electro mechanical system (MEMS), and performs adaptive beamforming to estimate the channel. In \cite{zhao2018channel}, the flight data tracking scheme is proposed using unscented Kalman filter (UKF) to calculate position, velocity, and direction cosine matrix (DCM). In \cite{fu2018multi}, the author offers a target tracking method based on information consensus for multiple UAVs. In all the aforementioned literatures, the beam tracking solutions are provided by integrating the sensor information and the beamforming signal.

This paper considers a beam tracking scheme using the flight sensor data such as GPS and INS signals and the beamforming signal, which can persistently track the signal transmitted from the UAV node to the ground station (GS) node.
To make use of those two different signals, we need to adjust these signals since they operate in a different mode, such as signaling, time periods, etc. We here point out that the GPS signal needs some correction to compensate for an error of about $5\sim15$ [m] and the GPS signal period is longer than that of general communication signals \cite{pixhawk4}.
Therefore, we propose a remedy to reduce the GPS signal's errors and enhance the limited 3D beamforming gain due to the sporadic GPS signal by using the beamforming signal.

The proposed scheme first designs the beamforming weight based on the measured GPS signal. It then updates the position with the beamforming signal, maintaining continuous high beamforming gain even in the absence of the GPS signal.
Moreover, we propose to employ Gaussian process regression (GPR) \cite{rasmussen2003gaussian,nguyen2009model,jang2020multi} for channel estimation. With the help of the flight sensor, the proposed scheme obtains the optimal beamforming weight using GPR. Note that GPR is a machine learning method to predict the outputs for unseen inputs based on the correlations of training inputs. We can employ GPR for a hybrid beamformer, including analog and digital beamformers and show that the proposed scheme for analog/hybrid beamforming outperforms the conventional adaptive gradient descent beamforming schemes presented in \cite{zhao2018beam,fakharzadeh2009fast}. We also show that our proposed scheme can predict beamspace channel information without additional measurements and can overcome the angular resolution limits of the phase shifter. Furthermore, we show from simulation results that the mean absolute error (MAE) as the performance metric of the channel estimator can be evolved to calculate beamforming performance in the data transmission block.

We here summarize our three main contributions to the UAV beam tracking scheme.
\begin{itemize}
    \item We propose a flight sensor data, such as GPS and INS, and beamforming signal based integrated UAV tracking scheme. Considering different specifications of GPS, INS, and the communication signal, we design a fully integrated framework, which can track the target continuously with high-accuracy even during the absence of a flight signal.
    \item We propose a novel GPR based channel estimation scheme for a hybrid and analog beamformer. The hybrid beamforming based proposed scheme can estimate the channel with faster convergence than conventional adaptive beamforming schemes. Moreover, the analog beamforming based proposed scheme shows improved performance than the codebook based beamforming scheme.  
    \item Our simulation results demonstrate that the 3D beamforming gain and spectral efficiency (SE) can be derived in terms of mean absolute error (MAE) of angle estimation. Therefore, we can calculate the beamforming performance metrics in the data transmission block in advance through the accuracy information of angle estimation for the pilot signals.
\end{itemize}

The rest of the paper is organized as follows. The system and channel models are explained under consideration in Sec. \ref{s:sysmodel}. In Sec. \ref{s:method}, we propose the flight sensor data and beamforming signal based integrated UAV tracking scheme.  The designing of the beamforming matrix by flight sensor data is presented in Sec. \ref{ss:step1}, and Sec. \ref{ss:step2} shows the final position update using GPR based channel estimation method. In Sec. \ref{sc:metric}, it can be shown that the beamforming performance metrics can be derived from the MAE of channel estimation.
We provide simulation results in Sec. \ref{s:simulation}, then present the concluding remarks in Sec. \ref{s:conclusion}.

\section{System Model} \label{s:sysmodel}
\begin{figure}[htbp]
    \centering
    \includegraphics[width=0.45\textwidth]{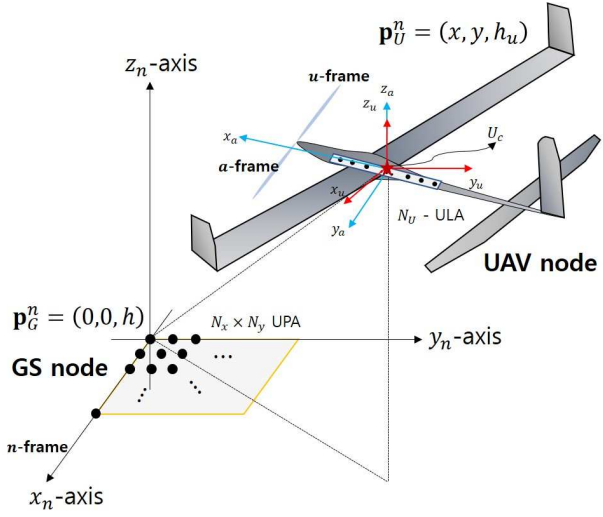}
    \caption{System model of UAV-aided communication system, where the GS node is equipped with a UPA of $N_x\times N_y$, while the UAV node has a $N_u$-ULA.}
    \label{fig:sys}
\end{figure}
We consider an integrated sensor aided UAV communication system depicted in Fig. \ref{fig:scenario}. The ground station (GS) node serves the UAV node in the A2G network with 3D beamforming gain. The GS node is equipped with a uniform planar array (UPA) of $N_x \times N_y$, which is on the $xy$-plane in the Cartesian coordinate as shown in Fig. \ref{fig:sys}, while the target UAV node has a uniform linear array (ULA) with $N_u$ number of antennas.
We assume that the GS node is located at the origin at the height of $h$, and the reference antenna element of UPA is positioned on the point at $\mathbf{p}_G^n=(0,0,h)$. We also assume that the antenna spacing is all equal to $d=\lambda/2$.

\begin{figure}
    \centering
    \includegraphics[width=0.4\textwidth]{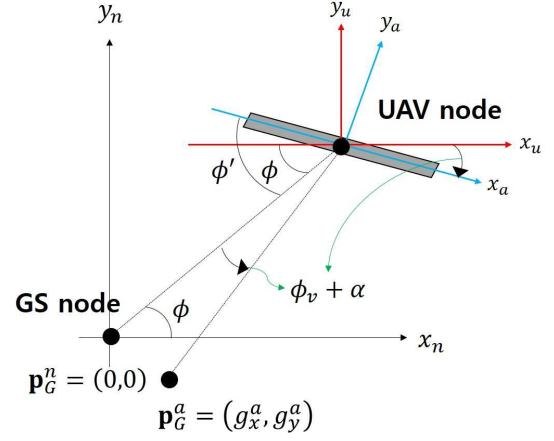}
    \caption{The definition of spatial angle of the channel.}
    \label{fig:coordinate}
\end{figure}

Most UAV nodes, equipped with sensors, can have access to flight-related data such as GPS, INS, and an on-board camera for navigation and self-positioning\cite{kim2003real,oh2010multisensor,angelino2012uav}. Especially, GPS provides a local position information $\mathbf{p}_U^n$ and the GS node receives the flight information message periodically for safety concerns. Moreover, we can obtain the attitude status of the UAV node from INS such as yaw $\alpha$, pitch $\beta$, and roll $\gamma$ of Euler angles. 
As in Fig. \ref{fig:coordinate}, $n$-frame represents a geodetic coordinate frame, whose origin is the center of gravity at the GS node with its three axes $x_n$, $y_n$, and $z_n$. While $u$-frame denotes a geodetic coordinate frame whose origin is at the center of the UAV node, and $a$-frame is the ULA coordinate frame at the UAV node.
The $u$-frame is parallel to the $n$-frame and the origin of the $a$-frame is the same as the center of the $u$-frame given as $U_c=(x,y,h_u)$ with its $x_a$-axis paralleled to the heading direction of the UAV node, which is rotated by $\phi_v$ from the $x_u$-axis, assuming here that $N_u$-ULA is aligned with the forward of the UAV. Its $z_a$-axis is identical with the $z_u$-axis. In Fig. \ref{fig:coordinate}, we show the projection of the UAV location on the $xy$-plane, where the azimuth arrival angle from the view of the GS node and the departure angle from the $a$-frame at the UAV node are denoted as $\phi$ and $\phi^\prime$ ($=\phi+\phi_v$), respectively.
In the same manner, the attitude yaw angle $\alpha$ makes the $a$-frame rotate on the $xy$-plane. Considering the attitude status, the azimuth angle of the UAV node can be written as $\phi^\prime=\phi+\phi_v+\alpha$. We herein denote the heading angle as $\phi_v+\alpha$. The pitch and roll angles can be negligible because the UAV node flies at the same height and we consider the UAV node equipped with ULA \cite{yang2019beam}. We point out that the pitch angle rotating on the $y$-axis is trivial in ULA since the beamforming power of ULA with a fixed azimuth angle is the same at any elevation angle. 
Let $\mathbf{p}_G^i$ denotes the GS node position at the $i$-frame and $\mathbf{p}_U^i$ denotes the the UAV node position at the $i$-frame, where $i$ represents $n$, $a$, and $u$.

The spatial arrival angle of the GS node on the $x$-axis and $y$-axis for the position at the $n$-frame can be defined, respectively, as
\begin{equation} \label{eq:uv_n}
    \begin{aligned}
    u&=\frac{x}{\sqrt{x^2+y^2+(h_u-h)^2}}, \\
    v&=\frac{y}{\sqrt{x^2+y^2+(h_u-h)^2}},
    \end{aligned}
\end{equation}
where (\ref{eq:uv_n}) can also be expressed as $\cos{\phi}\sin{\theta}$ and $\sin{\phi}\sin{\theta}$, where $\phi$ and $\theta$ denote azimuth and elevation angle, respectively. The corresponding spatial departure angle of the UAV node on the $x_a$-axis can be defined as
\begin{equation} \label{eq:uv_a}
    \begin{aligned}
    u_a=\frac{g_x^a}{\sqrt{(g_x^a)^2+(g_y^a)^2}}=\cos{\phi'},
    \end{aligned}
\end{equation}
where $(g_x^a,g_y^a)$ denotes the transformed position of the GS node in respect of the $a$-frame, which can be calculated by the coordinate transformation matrix from the $u$-frame to the $a$-frame, as 
\begin{equation} \label{eq:coord_trans}
    \mathbf{C}_u^a = \mathbf{T}_3(\gamma)\mathbf{T}_2(\beta)\mathbf{T}_1(\alpha).
\end{equation}
In (\ref{eq:coord_trans}), $\mathbf{T}_1(\alpha)$ represents the spatial rotation of yaw, $\alpha$ around the $z_u$ axis as 
\begin{equation}
    \mathbf{T}_1(\alpha)=\begin{bmatrix}\cos{\alpha}&\sin{\alpha}& 0\\-\sin{\alpha}&\cos{\alpha}&0\\0&0&1
\end{bmatrix},
\end{equation}
and $\mathbf{T}_2(\beta)$ represents rotating pitch, $\beta$ around the $y_u$ axis as 
\begin{equation}
    \mathbf{T}_2(\beta)=\begin{bmatrix}\cos{\beta}&0&-\sin{\beta}\\0&1&0\\\sin{\beta}&0&\cos{\beta}\end{bmatrix},
\end{equation}
and $\mathbf{T}_3(\gamma)$ represents rotating roll, $\gamma$ around $x_u$ axis as 
\begin{equation}
    \mathbf{T}_3(\gamma)=\begin{bmatrix}1&0&0\\0&\cos{\gamma}&\sin{\gamma}\\0&-\sin{\gamma}&\cos{\gamma}\end{bmatrix}.
\end{equation}
The transformed origin of the GS node in respect of the $a$-frame, $(g_x^a,g_y^a)$ can be obtained from $\mathbf{C}_u^a\cdot\mathbf{p}_G^u=\mathbf{C}_u^a\cdot(-\mathbf{p}_U^n)$.

\subsection{Channel Model}
We assume that the pilot signal is transmitted periodically to the GS node in every communication block.
At the $k$th communication block, the UAV node position at the $n$-frame is denoted as $\mathbf{p}_U^n(k)=\begin{bmatrix}x(k)&y(k)&h_u\end{bmatrix}^T$ with a constant height, $h_u$. The velocity of UAV node can be calculated by differentiating the local position denoted as 
$\mathbf{v}_u(k)=\begin{bmatrix}v_x(k)&v_y(k)&0\end{bmatrix}^T=\begin{bmatrix}v_u(k)\cos{\phi_{v}(k)}& v_u(k)\sin{\phi_{v}(k)}&0\end{bmatrix}^T$, where $\phi_v(k)$ is the heading angle of UAV node on the $xy$-plane as depicted in Fig. \ref{fig:coordinate}.
Following the first-order Gaussian-Markov model \cite{ma2019wideband}, we can write the magnitude and direction of velocity as
\begin{equation}
    \begin{aligned}
    v_u(k+1) &= \rho v_u(k) + \epsilon_v(k), \\
    \phi_{v}(k+1) &= \rho \phi_{v}(k) + \epsilon_d(k),
    \end{aligned}
\end{equation}
where $\rho$ is the correlation coefficient and the process noises, $\epsilon_v$ and $\epsilon_d$, follow
Gaussian distribution as $\epsilon_v, \epsilon_d \sim \mathcal{CN}(0,(1-\rho^2/2))$. 
Assuming that the UAV node flies at the constant velocity for the time duration of a communication block, we can represent the position transition model at the $n$-frame as
\begin{equation}
    \mathbf{p}_U^n(k+1) = \mathbf{p}_U^n(k)+T_p\mathbf{v}_u(k),
\end{equation}
where $T_p$ denotes the time length of a communication block.

Based on the millimeter-wave channel model, the channel on the $k$th communication block can be written as
\begin{equation} \label{eq:ch}
\begin{aligned}
    \mathbf{H}(k) =& \xi(k) \mathbf{a}_g\left(u(k),v(k)\right)\mathbf{a}^H_u\left(u_{a}(k)\right) \\&+ \sum_{i=1}^{L}\xi_{i}(t)\mathbf{a}_g\left(u_i(k),v_i(k)\right)\mathbf{a}^H_u\left(u_{a,i}(k)\right),
    \end{aligned}
\end{equation}
where $\xi(k)$ denotes the gain embracing antenna gain $G$, path loss $D^{-\beta_c}$, and channel complex gain $\mu$ as $\xi(k) = \frac{G\mu(k)}{D(k)^{\beta_c}}$, and $\beta_c$ is a path loss exponent. In (\ref{eq:ch}), $\mathbf{a}_g$ and $\mathbf{a}_u$ denote the array response vector of the GS and UAV nodes, respectively.
In (\ref{eq:ch}), the first term is a line-of-sight (LOS) path and the remaining term is the summation of non-LOS (NLOS) paths. Note that in the aerial network, the high probability of LOS path is often assumed \cite{zhao2018channel}. Therefore, we consider that the NLOS path can be negligible, then
the normalized dominant LOS channel $\bar{\mathbf{H}}(k)$ can be defined as 
\begin{equation} \label{eq:h_los}
    \bar{\mathbf{H}}(k)= \mu(k)\mathbf{a}_g\left(u(k),v(k)\right)\mathbf{a}^H_u\left(u_{a}(k)\right),
\end{equation}
where $u(k)$ and $v(k)$ can be modeled as in (\ref{eq:uv_n}), and
$u_{a}(k)$ denotes the spatial departure angle of the UAV node relative to the $a$-frame in (\ref{eq:uv_a}).
The array response vector of the GS node can be modeled as
\begin{equation}
\mathbf{a}_g(u(k),v(k))=\mathbf{a}_x(u(k)) \otimes, \mathbf{a}_y(v(k)),
\end{equation}
where the array response vector for the $x$-axis and the $y$-axis are defined, respectively, as
\begin{equation}
\begin{aligned}
    \mathbf{a}_x(u(k))&= \begin{bmatrix} 1\! & \!e^{-j\pi u(k)}\!& \!\cdots\! &\!e^{-j\pi(N_x-1)u(k)}\end{bmatrix}^T, \\
    \mathbf{a}_y(v(k))&= \begin{bmatrix} 1\! & \!e^{-j\pi v(k)}\!& \!\cdots\! &\!e^{-j\pi(N_y-1)v(k)}\end{bmatrix}^T.
\end{aligned}
\end{equation}
The array response vector of the UAV node can also be expressed as
\begin{equation}
    \mathbf{a}_u(u_a(k))= \begin{bmatrix} 1\! & \!e^{-j\pi u_a(k)}\!& \!\cdots\! &\!e^{-j\pi(N_u-1)u_a(k)}\end{bmatrix}^T.
\end{equation}

\subsection{3D Beamforming}
Considering the hybrid beamforming systems at the GS node, the received signal at the $k$th communication block can be written as
\begin{equation} \label{eq:rxsig}
\mathbf{r}(k)=\mathbf{W}^H(k)\bar{\mathbf{H}}(k)\mathbf{f}(k)\mathbf{s}(k)+\mathbf{W}^H(k)\mathbf{n}(k),
\end{equation}
where $\mathbf{W}(k)$ is the beamforming matrix, the beamforming precoding vector $\mathbf{f}(k)$ can be directly designed by the position and attitude information of the UAV node itself, which is discussed in detail in the next section, and the noise vector $\mathbf{n}(k)$ denotes complex Gaussian noise at the GS node with $\mathbf{n}\sim \mathcal{CN}(0,\sigma_n^2 \mathbf{I})$.
In (\ref{eq:rxsig}), we can write the beramforming matrix as $\mathbf{W}=\mathbf{W}_{RF}\times \mathbf{W}_{BB}$, so $\mathbf{W}_{RF}$ and $\mathbf{W}_{BB}$ need to be designed for the RF and the digital combinings, respectively.
The beamforming weight matrix $\mathbf{W}(k)$ is given as $\mathbf{W}(k)=\begin{bmatrix}\mathbf{w}_1(k)&\mathbf{w}_2(k)&\cdots&\mathbf{w}_G(k)\end{bmatrix}$, where $\mathbf{w}_i(k)$ can be denoted as  $\mathbf{w}_{x,j}\otimes\mathbf{w}_{y,k}$ for $i=j\cdot G_y+k$. We here define the length of codebook for the $y$-axis angular parameter as $G_y$. The array response vector of $\mathbf{w}_{x}$ and $\mathbf{w}_{y}$ can be defined as  $\frac{1}{\sqrt{N_m}}\begin{bmatrix}e^{-j\pi\psi}&,\cdots,&e^{-j\pi(N_m-1)\psi}\end{bmatrix}^T$ with $m=x,y$ and certain phase $\psi$.

\section{Proposed Flight Sensor and Beamforming Signal based Integrated Tracking Scheme} \label{s:method}
\begin{figure}[htbp]  
    \centering
    \includegraphics[width=0.45\textwidth]{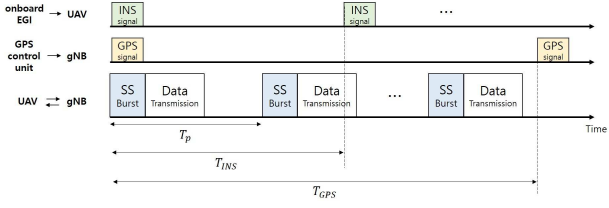}
    \caption{Three different signal frame models including on-board EGI signal at the UAV node, GPS control unit signals at the GS node, and the communication signals between the two nodes for the beam tracking.}
    \label{fig:sig}
\end{figure}

In general, the communication system can utilize flight sensor data to estimate channel efficiently. However, it is challenging to utilize it continuously since the period of the sensor message signal is usually longer than that of the communication signals. For example, the period of sensor in GPS-only system is about $50\sim100$ [ms] \cite{pixhawk4}, and the period of the INS signal is $20$ [ms] \cite{pixhawk4}, which are the results even with higher sampling data by calibration processing than raw data. On the other hand, the period of the 5G NR pilot frame as an example is $0.3\sim40$ [ms] \cite{3gpp.38.211}. In 5G NR standards, initial access needs a synchronization signal (SS) burst set to sweep the beams, and the tracking procedure needs SS and CSI-RS blocks to manage continuous beam tracking. \cite{giordani2018tutorial}. 

Let us consider that the GS node can obtain the position data from the GPS control unit on the ground every $T_{GPS}$. Note that GPS has a relatively large measurement error of about $5\sim15$m and it can be improved to about $2\sim5$m through calibration processing \cite{miura2015gps}. At the GS node, a set of beam candidates based on the measured GPS signal such as position and velocity is constructed at every $T_{p}$, which will be explained in the following section. The UAV node can have access to the position and attitude data from the on-board EGI at every $T_{INS}$, which can achieve higher accuracy on GPS data than the case that only the GPS signal is used. Moreover, the UAV node can easily obtain the precoding vector based on its attitude measurements with high accuracy \cite{unsal2012estimation,ahmed2016accurate,LN-100G}.

As in Fig. \ref{fig:sig}, the GS node receives the GPS signal at every $T_{GPS}$ and the UAV node performs precoding based on the EGI measurements at every $T_{INS}$, then the GS node estimates a channel at every $T_p$, $(T_p<T_{GPS})$ with the beamformed pilot signal. The proposed scheme can
maintain continuous high-accuracy tracking even when the GPS signal is absent.
In this section we describe our proposed UAV tracking scheme as 1) position prediction and beamforming matrix design based on the measurements of GPS and IMU, and 2) spatial angular parameter estimation, which can be applied for directional beamforming with high accuracy.

\subsection{Position Prediction using Flight Information} \label{ss:step1}
Assuming the UAV node flying with a constant velocity, let us define the predicted position, $\hat{\mathbf{p}}_U^n(k)^-$, at the $k$-th communication block as
\begin{equation} \label{eq:pred}
\begin{array}{ll}
\hat{\mathbf{p}}_U^n(k)^-\!= \Tilde{\mathbf{p}}_U^n(m), \ k=m,&\hbox{with GPS signal} \\
\hat{\mathbf{p}}_U^n(k)^-\!= \hat{\mathbf{p}}_U^n\!(k\!-\!1)\!+ T_p\Tilde{\mathbf{v}}_u(m),\!\! \ k\ne m,&\hbox{without GPS signal}
\end{array}
\end{equation}
Note that $\Tilde{(\cdot)}$ denotes the measurement value of the corresponding parameter, and $(\cdot)^-$ denotes the predicted transition parameter. In (\ref{eq:pred}), $k$ and $m$ denote the time index for the pilot signal and the GPS signal, respectively, and $\Tilde{\mathbf{p}}_U^n(m)$ denotes the position measured from the GPS signal with errors of variance, $\sigma_G^2$. The velocity value can be calculated as $\Tilde{v}(m)$ when the GPS signal exists, and the obtained value is used at every communication block as in (\ref{eq:pred}) until a new GPS signal is updated.

We also focus on the attitude of the UAV node's flight status. The heading direction and attitude status can change the spatial angle of the $a$-frame of the UAV node. Note that the UAV node obtains its position information based on the measurement from various sensors\cite{pixhawk4,LN-100G}. Using the measured data from the on-board EGI, which is quite accurate \cite{LN-100G}, we can obtain the spatial angle, $u_a$, at the UAV node as given in (\ref{eq:uv_a}).
The transformed relative position at the $a$-frame of the UAV node, applying the heading angle, can be obtained as
\begin{equation}
    \hat{\mathbf{p}}_{G}^a(k) = \mathbf{C}_u^a\left(-(\Tilde{\phi_v}(n)+\Tilde{\gamma}(n))\right)\Tilde{\mathbf{p}}_G^u(n),
\end{equation}
where $\Tilde{\mathbf{p}}_G^u(n)$ denotes a relative position of the GS node at the $u$-frame, and $n$ is the time index for the INS and GPS signals at the UAV node. We can also obtain the transformed spatial departure angle of the UAV node as 
\begin{equation}
    \hat{u}_a=\cos{\left(\hat{\phi}'(k)\right)}=\cos{\left(\tan{\left(\frac{\Tilde{y}_U^n(n)}{\Tilde{x}_U^n(n)}\right)}+\Tilde{\phi}_h(n)\right)},
\end{equation}
where $\Tilde{x}_U^n(n)$ and $\Tilde{y}_U^n(n)$ denote measured position data of the flight controller with errors of variance, $\sigma_{G,ins}^2$, and $\Tilde{\phi}_h(n)$ is the measurement of heading angle embracing the yaw angle, $\alpha$, with errors of variance as $\sigma_I^2$. We assume that the position and attitude measurements can be sampled with higher frequency as $T_{INS}$ in the on-board INS/GPS controller than in the GPS control unit in the ground. 
Then, the UAV node can generate the beam toward departure angle $\hat{u}_a$ with the flight information. The obtained beamforming precoding vector can be represented as
\begin{equation} \label{eq:precod}
    \mathbf{f}=\frac{1}{\sqrt{N_u}}\begin{bmatrix}1&e^{j\pi\hat{u}_a}&\cdots&e^{j\pi(N_u-1)\hat{u}_a}\end{bmatrix}^T.
\end{equation}
Note that the UAV node can accomplish the beam alignment without the downlink beam management procedure. The accuracy of the estimated phase $\hat{u}_a(k)$ in (19) is quite high by using both the INS and GPS signals \cite{ahmed2016accurate,LN-100G}.  

At the GS node, the power of the pilot signals can be measured by beamforming on the effective channel $\bar{\mathbf{H}}_{e}=\bar{\mathbf{H}}\mathbf{f}$ in the uplink channel. 
To determine the beamforming directions for the pilot signals, we exploit 
the the UAV node position $\Tilde{\mathbf{p}}_U^n$ measured from the GPS signal.
The obtained spatial angles from the GPS signal can be represented as
\begin{equation}
\begin{aligned} \label{eq:uv_err}
    \Tilde{u}(k)=\frac{\Tilde{x}(k)}{\sqrt{\Tilde{x}(k)^2+\Tilde{y}(k)^2+h^2}}=u(k)+\Delta u(k), \\
    \Tilde{v}(k)=\frac{\Tilde{y}(k)}{\sqrt{\Tilde{x}(k)^2+\Tilde{y}(k)^2+h^2}}=v(k)+\Delta v(k),
\end{aligned}
\end{equation}
where $\Tilde{x}(k)$ and $\Tilde{y}(k)$ are Gaussian random variable as $\mathcal{N}(x(k),\sigma_G^2)$ and $\mathcal{N}(y(k),\sigma_G^2)$, respectively, and $\Delta u(k)$ and $\Delta v(k)$ denote deviated spatial angles that occurred from the GPS errors.
We can determine the angle candidate set around the point $(\Tilde{u}(k),\Tilde{v}(k))$ as
\begin{equation} \label{eq:bfset_p}
\begin{aligned} 
    \mathcal{D}_s=\Bigg\{(u_{i},v_j),&u_i\in\left[\Tilde{u}-B_s:\Delta:\Tilde{u}+B_s\right],\\ &v_j\in\left[\Tilde{v}-B_s:\Delta:\Tilde{v}+B_s\right]\Bigg\},
\end{aligned}
\end{equation}
where $\Delta$ denotes the angular grid interval, and $B_s$ is the length of the defined codebook. 

\subsection{Position Update based on Channel Estimation using GPR} \label{ss:step2}
Gaussian Process Regression (GPR) is a prediction method in function-space view based on a non-parametric Bayesian model \cite{rasmussen2003gaussian}. GP is a combination of prior distribution and observations, connecting to posterior distribution over functions, which are randomly generated with zero-mean and kernel covariance.
The unknown prior functions are assumed to be zero-mean Gaussian distribution as $f(\mathbf{x})\sim \mathcal{GP}\left(0,k(\mathbf{x},\mathbf{x}')\right)$ with a kernel squared exponential (SqExp) function, $k(\mathbf{x},\mathbf{x}')$ defined as 
\begin{equation}
    k(\mathbf{x},\mathbf{x}')=\sigma_s^2 \exp{\left(-\frac{1}{2}(\mathbf{x}-\mathbf{x'})^T\Sigma^{-1}(\mathbf{x}-\mathbf{x'})^T\right)},
\end{equation}
where $\sigma_s^2$ denotes variance of $f(\mathbf{x})$, and $\Sigma$ is the length parameter that determines how fast the correlation between inputs decreases.

We make use of a set of the training input data in 2D angular space with grid points $\mathbf{X}=\{\mathbf{x}_i\}_{i=1}^{G_s}$, which contains neighborhood points around $(\Tilde{u}(k),\Tilde{v}(k))$ in (\ref{eq:bfset_p}), and denote $\mathbf{x}_i=\begin{bmatrix}u_i,v_i\end{bmatrix}^T$.
Note that the observed outputs of training set is the magnitude of
beamformed signals in (\ref{eq:rxsig}) as $\mathbf{y}_t=f(\mathbf{X})+\xi$, where $\xi$ is a measurement noise. Here, we can choose hyperparameters as $\theta=[\sigma_s,\Sigma]$ of kernel function directly from the training data. It is important to set the fitted kernel function in GPR. We can express the marginal likelihood function as
\begin{equation} \label{eq:likelihood}
    \mathcal{L}=\log{p(\mathbf{y}_t|\mathbf{X},\mathbf{\theta})}=-\frac{1}{2}\mathbf{y}_t^T \mathbf{K}^{-1}\mathbf{y}_t-\frac{1}{2}\log{\lvert \mathbf{K}\rvert}-\frac{G_s}{2}\log{2\pi},
\end{equation}
where $\mathbf{K}$ denotes the kernel matrix from the training set.
To select optimal hyperparameters maximizing the marginal likelihood, we use the partial derivatives of the marginal likelihood with regards to the hyperparameters as
\begin{equation}
    \frac{\partial \mathcal{L}}{\partial \mathbf{\theta}_j} =\frac{1}{2}\mathbf{y}_t \mathbf{K}^{-1}\frac{\partial \mathbf{K}}{\partial \mathbf{\theta}_j}\mathbf{K}^{-1}\mathbf{y}_t-\frac{1}{2}\mathrm{tr}\left(\mathbf{K}^{-1}\frac{\partial \mathbf{K}}{\partial \mathbf{\theta}_j}\right), j=1,2.
\end{equation}
Then, the prior joint distribution of the observed measurements and the function values at new sample $\mathbf{x}^*$ can be expressed as
\begin{equation} \label{eq:prior}
    \begin{bmatrix}\mathbf{y}_t\\\mathbf{f(\mathbf{x}^*)}\end{bmatrix}\sim \mathcal{N}\left(\mathbf{0},\begin{bmatrix}\mathbf{K}(\mathbf{X},\mathbf{X})+\sigma_n^2\mathbf{I}&\mathbf{K}(\mathbf{X},\mathbf{x}^*)\\\mathbf{K}(\mathbf{x}^*,\mathbf{X})&\mathbf{K}(\mathbf{x}^*,\mathbf{x}^*)\end{bmatrix}\right),
\end{equation}
where $\mathbf{K}(\mathbf{X},\mathbf{X})$ denotes the $G_s\times G_s$ covariance matrix evaluated over all training points, and $\mathbf{K}(\mathbf{X},\mathbf{x}^*)$ denotes $G_s\times 1$ vector calculated over at all pairs of the training set and test point. We can set new test input $\mathbf{x}^*$ as the new spatial angle value which is not contained in the training set.

The posterior distribution which is conditioned by training inputs, observed values, and new test set can be derived as
\begin{equation} \label{eq:gpr}
    \mathbf{f}^*|\mathbf{X},\mathbf{y}_t,\mathbf{x}^* \sim \mathcal{N}\left(\hat{\mathbf{f}}^*,\Sigma^*\right),
\end{equation}
where posterior mean and variance are defined, respectively, as
\begin{equation} \label{eq:posterior}
    \begin{aligned}
    \hat{\mathbf{f}}^*&=K(\mathbf{x}^*,\mathbf{X})\left(K(\mathbf{X},\mathbf{X})+\sigma_n^2\mathbf{I}\right)^{-1}\mathbf{y}_t, \\
    \Sigma^*&=K(\mathbf{x}^*,\mathbf{x}^*)-K(\mathbf{x}^*,\mathbf{X})\begin{bmatrix}K(\mathbf{X},\mathbf{X})+\sigma_n^2\mathbf{I}\end{bmatrix}^{-1}K(\mathbf{X},\mathbf{x}^*).
    \end{aligned}
\end{equation}
In (\ref{eq:posterior}), we denote the new test sample as $\mathbf{x}^*$ and it can also be extended to new test set as $\mathbf{X}^*$.

\begin{figure}[htbp]
    \centering
    \includegraphics[width=0.45\textwidth]{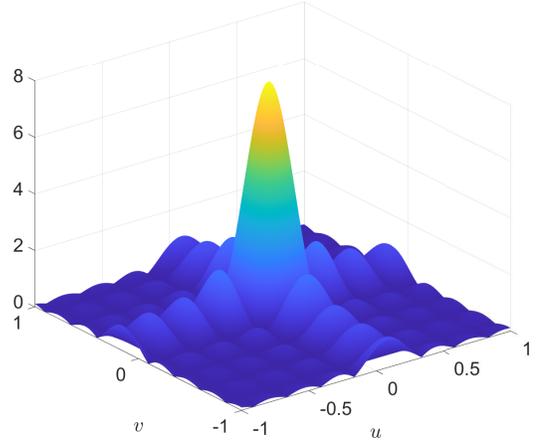}
    \caption{Beam space channel model with spatial angle parameters $u=\cos{\phi}\sin{\theta}=0.1504$ and $v=\sin{\phi}\sin{\theta}=0.0868$, where $\phi$ and $\theta$ are azimuth and elevation angle, respectively, for $8\times8$-UPA.}
    \label{fig:ch_bspace}
\end{figure}
In Fig. \ref{fig:ch_bspace}, we illustrate the beam space channel model, which represents coupling with spatial beam of the DFT matrix.
The angular values having the global maximum output are the channel parameters of $u$ and $v$. In this GPR based channel estimation, our objective is to find the spatial angle parameters $u$ and $v$.

\subsubsection{GPR based Hybrid Beamforming} \label{sc:hybrid}
The proposed channel estimation for the hybrid beamformer is similar to the adaptive beamforming. The objective of the adaptive beamforming is to find an optimal weight to provide the maximum signal power or null the interference \cite{gershman2010convex}. The adaptive beamformer updates the beamforming weight recursively to set the beam toward the desired direction. 

To find the optimal angle point on 2D angular space efficiently, we employ gradient descent method \cite{boyd2004convex} with GPR. We first select the input training set $\mathcal{D}_s$ in (\ref{eq:bfset_p}). Let us define $\Delta$ as $\frac{2\pi}{2^l}$ by the quantization bits of phase shifter, $l$, and we determine the length of the codebook $B_s$ as the null-to-null beamwidth of the main lobe, that is, $\frac{2}{N_x}=\frac{2}{N_y}$. We point out here that we only need to train the main lobe part to find the global maximum of the beam space channel. The range of training input $B_s$ can also be different for $u$ and $v$, respectively, if $N_x$ and $N_y$ are not the same. As a result, the input training set can be reduced by GPS data compared to searching the whole angular grid.

We can write the magnitude measurements of the beamformed signal at the GS node for determined directions $\mathcal{D}_s$ as
\begin{equation} \label{eq:rxpow}
    \mathbf{y}_t=\lvert\mathbf{r}\rvert=\lvert\mathbf{W}^H_{s}\bar{\mathbf{H}}\mathbf{f}\mathbf{s}+\mathbf{W}^H_{s}\mathbf{n}\rvert,
\end{equation}
where $\mathbf{W}_{s}$ is the corresponding beamforming matrix for $\mathcal{D}_s(k)$. 
Using (\ref{eq:gpr}), we have the derivatives of the predicted mean for new input $x^\ast$ as
\begin{equation} \label{eq:partial_f}
    g(\mathbf{x}^*)=\frac{\partial \hat{f}^*}{\partial \mathbf{x}^*}\Bigg|_{\mathbf{x}^*}=\frac{\partial \mathbf{k}^T(\mathbf{x}^*,\mathbf{X})}{\partial \mathbf{x}^*}\left(K(\mathbf{X},\mathbf{X})+\sigma_n^2\mathbf{I}\right)^{-1}\mathbf{y}_t.
\end{equation}
The channel model in beam space is uni-modal within the main robe region and becomes an approximated uni-modal shape in entire angular space as the number of antennas increases. The problem of getting stuck at local optima does not occur when the initial point is set in the neighborhood of global optimum, $u$ and $v$ as in Fig. \ref{fig:ch_bspace}. We can set the initial point as the particular input $\mathbf{x}$, whose corresponding measurement is the maximum among the measurements $\mathbf{y}_t$.

We can update the spatial angle, which can be employed to beamforming weight, as 
\begin{equation} \label{eq:update}
    \mathbf{x}^*_{t+1}=\mathbf{x}^*_t -\eta g(\mathbf{x}^*_t),
\end{equation}
where $\eta$ denotes the updating step size. 
By adding the measurements of the updated angle to the training set, the proposed scheme can estimate the output of channel function more accurately. 
Updating the estimated angle recursively is similar to the adaptive beamforming except the fact that the proposed scheme obtains channel information by GPR, which reduces the number of iteration compared to the adaptive beamforming scheme.
For example, the conventional schemes generally exploit the perturbation parameters \cite{zhao2018beam}. 
In the adaptive beamforming based on the gradient descent method \cite{fakharzadeh2009fast,zhao2018beam,zhao2014adaptive}, they obtain the derivatives by measuring the power for perturbed beamforming weight. 
We can show later on from our simulations that our proposed scheme effectively approaches the maximal point with less overhead and converges faster than the conventional adaptive based beamforming schemes. Moreover, it can be shown that the derivatives of the beam space can be obtained by GPR without beam steering toward the perturbed directions.

Utilizing the estimated $\hat{u}$ and $\hat{v}$ of $\mathbf{x}^*_{t+1}$ at the final iteration, we can update the predicted position $\hat{\mathbf{p}}_U^n(k)^-$ with high accuracy stabilizing the large errors of the GPS signal.
The obtained position $\hat{\mathbf{p}}_U^n(k)$ at the $k$th communication block can be derived as
\begin{equation}
\begin{aligned} \label{eq:pos_est}
\hat{x}(k) &= \frac{\hat{u}(k)h}{\sqrt{1-\left(\hat{u}(k)^2+\hat{v}(k)^2\right)}},\\
\hat{y}(k) &= \frac{\hat{v}(k)h}{\sqrt{1-\left(\hat{u}(k)^2+\hat{v}(k)^2\right)}}.
\end{aligned}
\end{equation}

\begin{algorithm}
\caption{GPR based Channel Estimation for Hybrid Beamforming} \label{algo1}
\begin{algorithmic} [1]
\STATE Measurements from GPS and IMU $\Tilde{\mathbf{p}}_U^n, \Tilde{\gamma}$
\STATE Prediction for position in (\ref{eq:pred})
\STATE Design a Precoding vector in (\ref{eq:precod})
\STATE Design a training input set in (\ref{eq:bfset_p})
\STATE Measure the beamformed signals with $\mathbf{W}$ in (\ref{eq:bfset_p}), (\ref{eq:rxpow})
\STATE Set the initial $\mathbf{x}^*$ among training set
\WHILE{$\lvert y_t^*-y_{t-1}^* \rvert<\epsilon$}
\STATE Measure the received signal magnitude $y_t^*$ for beamforming toward $\mathbf{x}^*_{t}$
\STATE Update the training set $\mathbf{X}\leftarrow \{\mathbf{X} \cup\mathbf{x}^*_{t}\}$
\STATE Predict the channel and gradient in (\ref{eq:gpr}) and (\ref{eq:partial_f})
\STATE Update the spatial angle $\mathbf{x}^*_{t+1} = \mathbf{x}^*_{t} - \eta g(\mathbf{x}^*_{t})$
\ENDWHILE
\STATE Estimate position $\hat{\mathbf{p}}_U^n(k)$ in (\ref{eq:pos_est})
\end{algorithmic}
\end{algorithm}

\subsubsection{GPR based Analog Beamforming}
The GPR based channel estimation method can be extended for the analog beamformer. In the beam training algorithms based on the codebook, the best codeword having maximum power measurement can be selected for the optimal beamforming weight in the codebook \cite{zhong2020novel}. The estimated angle is a quantized value constrained by phase resolution. However, the proposed scheme can derive the optimal off-grid angle from the predicted channel information. In Sec. \ref{sc:hybrid}, we update the training set $\mathbf{X}$ by beamforming toward the updated input as in Algorithm \ref{algo1}. Note that in the analog beamformer, we cannot generate the beam toward the desired off-grid direction since it is outside the codebook. Thus, we only use the training measurements from $\mathcal{D}_s$ without updating the training set. With the initially defined training set, we calculate the channel function and partial derivatives in (\ref{eq:gpr}) and (\ref{eq:partial_f}). The proposed scheme for the analog beamformer can update the estimated angle at each iteration toward near the optimal point of the corresponding channel.

\begin{algorithm}
\caption{GPR based Channel Estimation for Analog Beamforming} \label{algo2}
\begin{algorithmic}[2] 
\STATE $1-6$ lines in Algorithm \ref{algo1}
\WHILE{$\lvert f^*_t-f^*_{t-1} \rvert<\epsilon$}
\STATE Predict the channel $f^*_t$ in (\ref{eq:gpr}) and (\ref{eq:partial_f})
\STATE Update the spatial angle $\mathbf{x}^*_{t+1} = \mathbf{x}^*_t - \eta g(\mathbf{x}^*_{t})$
\ENDWHILE
\STATE Estimate position $\hat{\mathbf{p}}_U^n(k)$ in (\ref{eq:pos_est})
\end{algorithmic}
\end{algorithm}

It can be observed that the main computational complexity comes from the matrix inversion in (\ref{eq:prior}) and (\ref{eq:posterior}) in the training and the prediction step. The inversion term of the training data can be reused in the prediction, so we only consider the complexity of the training step. The inversion complexity of the covariance matrix is $\mathcal{O}(G_s^3)$, where $G_s$ the number of the training set. 
Adopting the sparse pseudo-input GP \cite{snelson2006sparse}, we can reduce the complexity to $\mathcal{O}(G_s)$ for the covariance matrix inversion of the training set.

\section{Effect of Angular Estimation Errors on Beamforming performance} \label{sc:metric}
As shown in the previous section, the GS node estimates the position of the UAV node in a pilot block. The GS node transmits/receives information bearing data to/from the UAV node in the data transmission block based on the estimated information in the pilot block. As a result, the errors in estimating the angular parameter can affect communication performance in the data transmission block. We here adopt the mean absolute error (MAE) of the angular estimation as a performance metric.

Let us write the obtained beamforming weight vector $\mathbf{w}(\hat{u}_d,\hat{v}_d)$ for the data transmission as
\begin{equation}
\begin{aligned}
    \mathbf{w}(\hat{u}_d,\hat{v}&_d)=\mathbf{w}_x(\hat{u}_d) \otimes \mathbf{w}_y(\hat{v}_d), \\
    \mathbf{w}_x(\hat{u}_d)&= \frac{1}{\sqrt{N_x}} \begin{bmatrix} 1\! & \!e^{-j\pi\hat{u}_d}\!& \!\cdots\! &\!e^{-j\pi(N_x-1)\hat{u}_d}\end{bmatrix}^T, \\
    \mathbf{w}_y(\hat{v}_d)&= \frac{1}{\sqrt{N_y}} \begin{bmatrix} 1\! & \!e^{-j\pi\hat{v}_d}\!& \!\cdots\! &\!e^{-j\pi(N_y-1)\hat{v}_d}\end{bmatrix}^T,
\end{aligned}
\end{equation}
where $\hat{u}_d$ and $\hat{v}_d$ denote the estimated angular values on the $x$-axis and $y$-axis at the final iteration in the pilot block, respectively. 

The received signal in the transmission block can be expressed as
\begin{equation} \label{eq:rx_sig}
    r = \mathbf{w}^H \bar{\mathbf{H}}_e s+ \mathbf{w}^H \mathbf{n},
\end{equation}
where $\bar{\mathbf{H}}_e=\sqrt{N_u}\mathbf{a}_g(u,v)$ is the effective channel and $\mathbf{n}$ denotes the complex Gaussian noise with the variance as $\sigma_n^2$.
Note that we can rewrite (\ref{eq:rx_sig}) as $r=p\left(\hat{u}_d,\hat{v}_d\right)+\omega$, where $\omega$ denotes the combining noise in the baseband. The magnitude of $p\left(\hat{u}_d,\hat{v}_d\right)$, that is, the beamforming gain can be written as 
\begin{equation}\label{eq:p_bar}
    \bar{p}\left(\hat{u}_d,\hat{v}_d\right) = \sqrt{N_u}\frac{\sin{\left(\frac{\pi N_x}{2}(u-\hat{u}_d)\right)}}{\sqrt{N_x}\sin{\left(\frac{\pi}{2}(u-\hat{u}_d)\right)}}
    \frac{\sin{\left(\frac{\pi N_y}{2}(v-\hat{v}_d)\right)}}{\sqrt{N_y}\sin{\left(\frac{\pi}{2}(v-\hat{v}_d)\right)}}.
\end{equation}
Employing MAE defined as $\delta=\mathbb{E}\left[\lvert\mathbf{x}-\hat{\mathbf{x}}\rvert\right]$, the expectation of (\ref{eq:p_bar}) can be rewritten in terms of $\delta$ as
\begin{equation}
\begin{aligned}
    \mathbb{E}&\left[\bar{p}\left(\hat{u}_d,\hat{v}_d\right)\right]\simeq \hat{\bar{p}}\left(\delta\right)\\
    &=\sqrt{N_u}\frac{\sin{\left(\frac{\pi N_x}{2}\delta\right)}}{\sqrt{N_x}\sin{\left(\frac{\pi}{2}\delta\right)}}
     \frac{\sin{\left(\frac{\pi N_y}{2}\delta\right)}}{\sqrt{N_y}\sin{\left(\frac{\pi}{2}\delta\right)}}, \\
     & -\frac{2}{N_x}\le\delta\le \frac{2}{N_x}, -\frac{2}{N_y}\le\delta\le \frac{2}{N_y},
    \end{aligned}
\end{equation}
where the estimated beamforming gain is only feasible on the main lobe. The 3D beamforming gain model has the same magnitude at a point with the same radius $\delta$ from the angular values $u$ and $v$ of the channel as depicted in Fig. \ref{fig:ch_bspace}.
We can also express the expectation of signal-to-noise ratio (SNR) of the beamformed signal as
\begin{equation}
    \mathrm{SNR}_o=\mathbb{E}\left[\frac{E_s \bar{p}\left(\hat{u}_d,\hat{v}_d  \right)^2}{\sigma_n^2\lVert\mathbf{w}\rVert^2}\right]\simeq\frac{E_s \hat{\bar{p}}\left(\delta\right)^2}{\sigma_n^2\lVert\mathbf{w}\rVert^2}
\end{equation}
where $E_s$ denotes average symbol energy.
The corresponding spectral efficiency (SE) can be written as
\begin{equation} \label{eq:se}
    \mathrm{SE}\simeq\mathrm{log}_2\left(\left\vert{1+\frac{E_s \hat{\bar{p}}\left(\delta\right)^2}{\sigma_n^2\lVert\mathbf{w}\rVert^2}}\right\vert\right) \mathrm{[bps/Hz]},
\end{equation}
where we need to consider the effect of estimation error, $\delta$. If we have prior information for the accuracy of the channel estimator such as MAE, we can estimate the beamforming gain and achieved SE in the transmission block in advance.

\section{Numerical Evaluation} \label{s:simulation}
\begin{figure*}[ht]
\begin{multicols}{2}
    \centering
    \subcaptionbox{Position tracking with relatively small errors as $\sigma_G=2$ [m] and $\mathrm{SNR}=20$ [dB]. }{\includegraphics[width=0.4\textwidth]{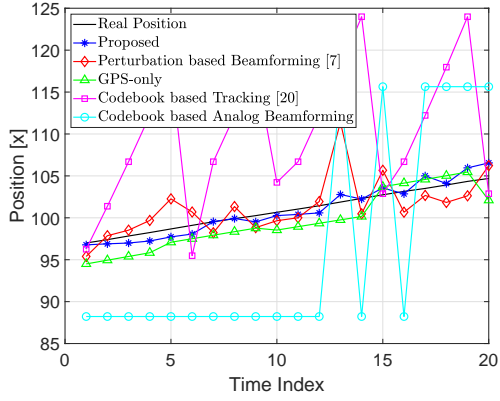}}
    \centering
    \subcaptionbox{Position tracking with relatively large errors as $\sigma_G=5$ [m] and $\mathrm{SNR}=10$ [dB].}{\includegraphics[width=0.4\textwidth]{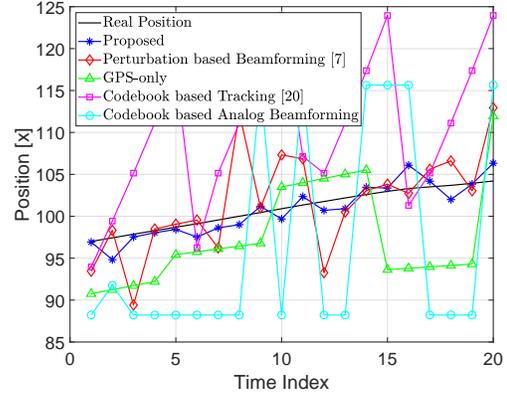}}
\end{multicols}
\caption{Position Tracking for proposed scheme versus other schemes.}
    \label{fig:pos_tracking}
\end{figure*}
\begin{figure*}[ht!]
\begin{multicols}{2}
    \centering
    \includegraphics[width=0.4\textwidth]{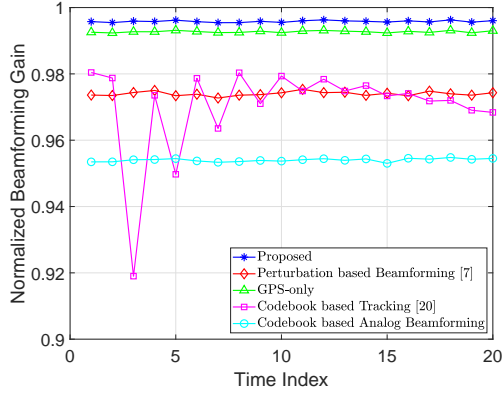}\par 
    \subcaption{Normalized beamforming gain for proposed scheme versus \\ other schemes with relatively high sensing accuracy as $\sigma_G=2$ [m],\\ $\sigma_{G,ins}=1$ [m], and $\sigma_I=0.01^\circ$.}
    \centering
    \includegraphics[width=0.4\textwidth]{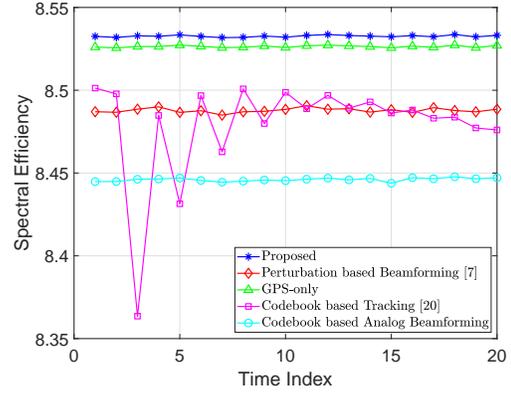}\par
    \subcaption{Spectral efficiency for proposed scheme versus other schemes\\ with relatively high sensing accuracy as $\sigma_G=2$ [m], $\sigma_{G,ins}=1$ [m], and $\sigma_I=0.01^\circ$.}
    \end{multicols}
\begin{multicols}{2}
\centering
    \includegraphics[width=0.4\textwidth]{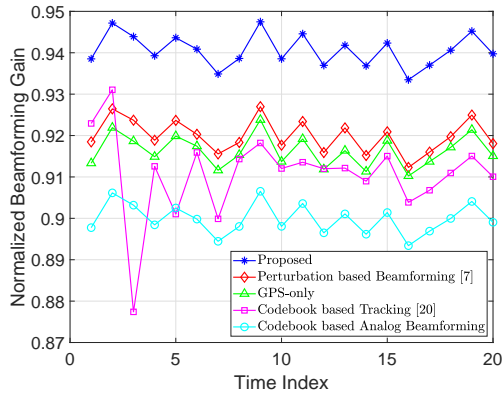}\par
   \subcaption{Normalized beamforming gain for proposed scheme versus \\ other schemes with relatively poor sensing accuracy as $\sigma_G=5$ [m],\\ $\sigma_{G,ins}=5$ [m], and $\sigma_I=0.05^\circ$.}
   \centering 
    \includegraphics[width=0.4\textwidth]{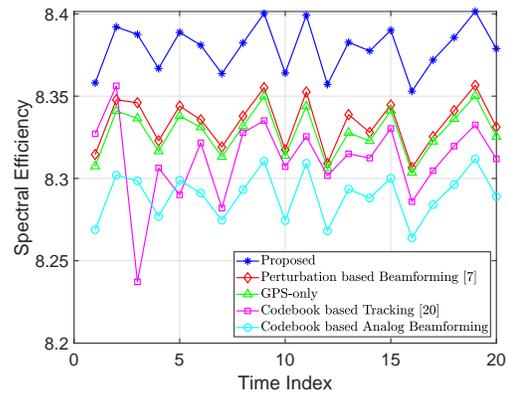}\par
    \subcaption{Spectral efficiency for proposed scheme versus other schemes\\ with relatively poor sensing accuracy as $\sigma_G=5$ [m], $\sigma_{G,ins}=5$ [m], and $\sigma_I=0.05^\circ$.}
\end{multicols}
\caption{Beamforming performances for proposed scheme versus other schemes.}
 \label{fig:performance_tracking}
\end{figure*}
In this section, we present the simulation results of the proposed position tracking and channel estimation schemes and compare them with other conventional schemes.

We set the location of the GS node as $(0,0,25)$ [m] as an example and assume that the position of the UAV node is random variable following the uniform distribution $x,y\in[10,100]$ with the height of $200$ [m]. We also set the number of antennas at the GS and the UAV nodes as $8$, that is, $N_x=N_y=N_u=8$ and the signal period as $T_{GPS}=50$ [ms], $T_{INS}=20$ [ms], and $T_p=10$ [ms], respectively \cite{pixhawk4,giordani2018tutorial}. We consider UAV with limited speed between $40$ [km/hr] and $160$ [km/hr], which is a normal operating speed of UAV \cite{huang20203d,miao2020lightweight}. For the analog beamforming, we assume $6$ bit phase resolution as an example, which will be explained detailed later on.

We provide the results of different beam tracking schemes to compare the performance. The perturbation based adaptive beamforming uses perturbation \cite {fakharzadeh2009fast,zhao2018beam,zhao2014adaptive} to find the optimal beamforming weight for maximum beamforming gain. The GPS-only scheme only uses GPS measurements when GPS signal comes in. The codebook based tracking \cite{huang20203d} uses the beamformed signal to calculate the speed of the UAV. The codebook based analog beamforming selects the angle candidate with the maximum power, assuming the $6$-bit resolution of the phase shifter.

\subsection{Position Tracking}
In Figs. \ref{fig:pos_tracking} and \ref{fig:performance_tracking}, we present the progress of the movement tracking and performance metrics over time index $k$ from $1$ to $20$. The time interval between $k$ and $k+1$ is set as the period of the pilot signal, $T_p$. The GS node receives the GPS signal at every $5$ time index, and the UAV node receives GPS/INS signal at every $2$ time index. We also assume that the UAV node performs precoding with the aid of EGI measurements, then the GS node estimates the angular value of effective channel and the position of the UAV node.

Fig. \ref{fig:pos_tracking} shows the position tracking in real-time compared to other tracking schemes. We set the measurement error of the GPS control unit with standard deviation, $\sigma_G$ as $2$ [m] and $5$ [m], respectively, and also set the value of SNR as $20$ [dB] and $10$ [dB], respectively. The UAV node is assumed to obtain the ideal position data of itself as $\sigma_{G,ins}=0$ and the attitude data with error whose standard deviation is $\sigma_I=0.01^\circ$ \cite{LN-100G}. We can see from Fig. \ref{fig:pos_tracking} that the GPS-only scheme is susceptible to the errors of the GPS signal. For example, if the error of the $15$th measurement is large, it can be seen that the large deviation from the real position is mostly maintained during the time from the $15$th period to the $19$th period.

\begin{figure}[htbp]
    \centering
    \subcaptionbox{Mean squared error of angular parameter estimation as $\hat{u}-u$ and $\hat{v}-v$, where $u=\cos{\phi}\sin{\theta}$ and $v=\sin{\phi}\sin{\theta}$ are the phase difference on the $x$-axis and $y$-axis based on $8\times8$-UPA, respectively. }{\includegraphics[width=0.48\textwidth]{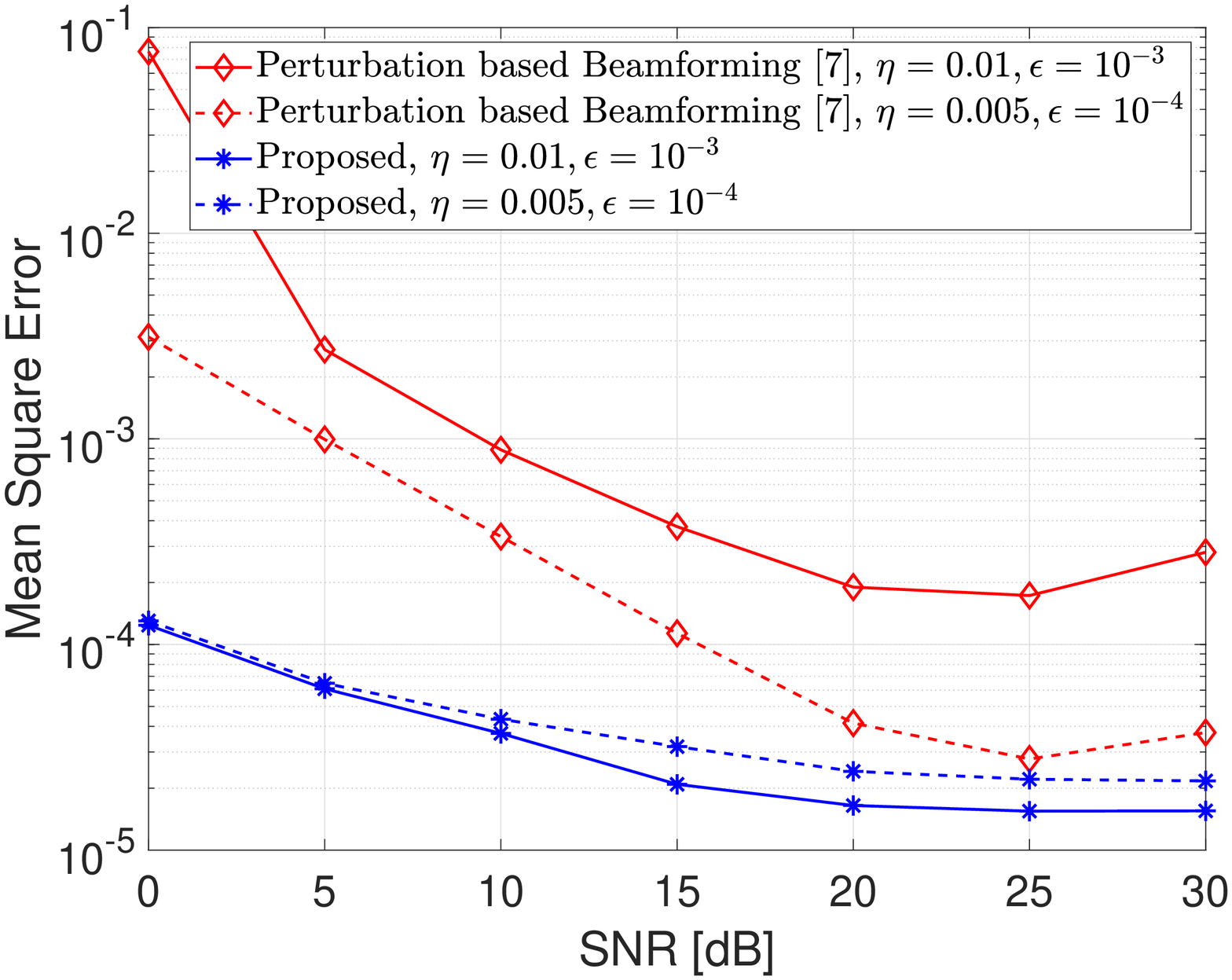}}
    \centering
    \subcaptionbox{The number of iterations for convergence.}{\includegraphics[width=0.48\textwidth]{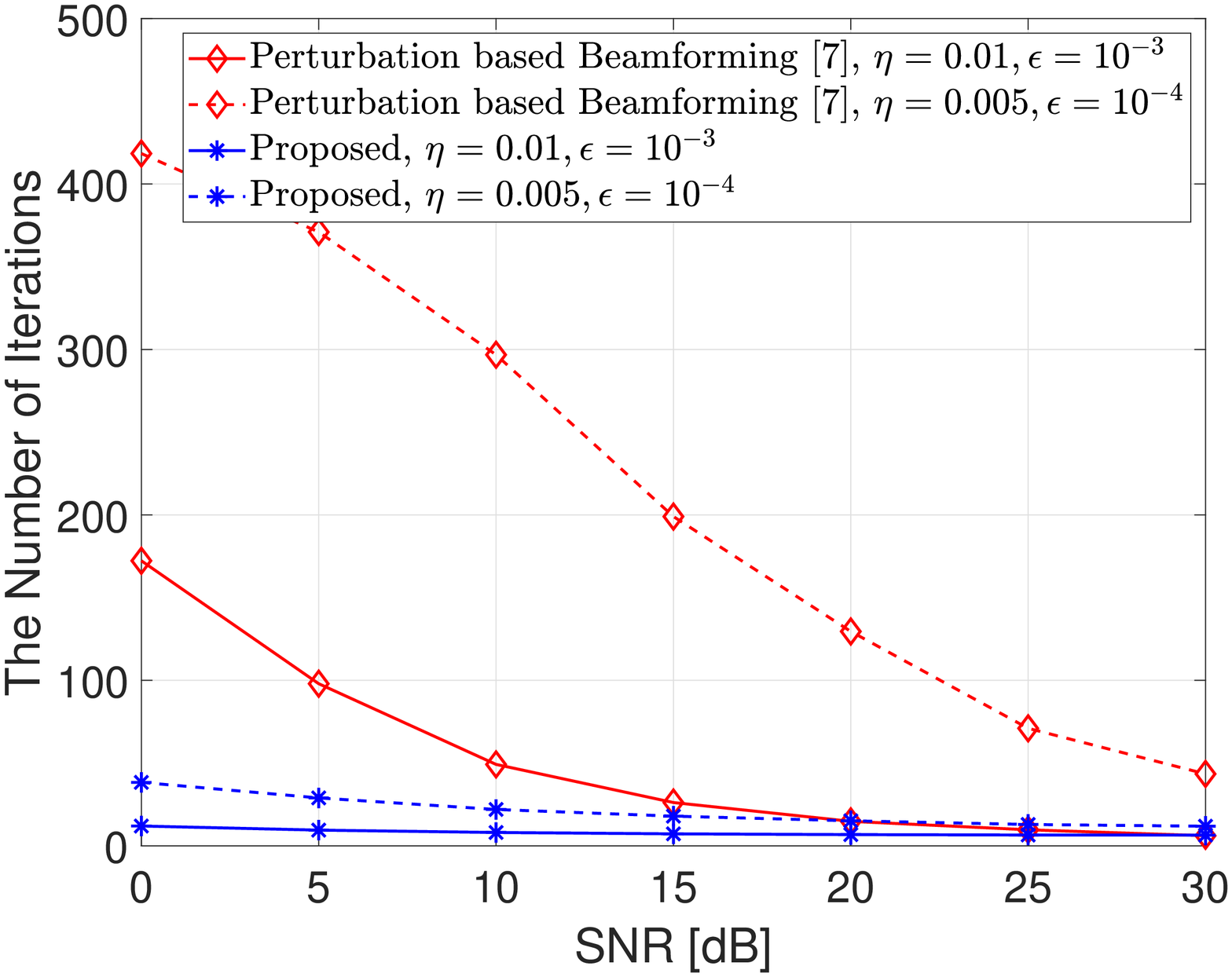}}
    \centering
    \subcaptionbox{Spectral Efficiency of directional beamforming for the estimated spatial angle as $\hat{u}$ and $\hat{v}$ based on $8\times8$-UPA. }{\includegraphics[width=0.48\textwidth]{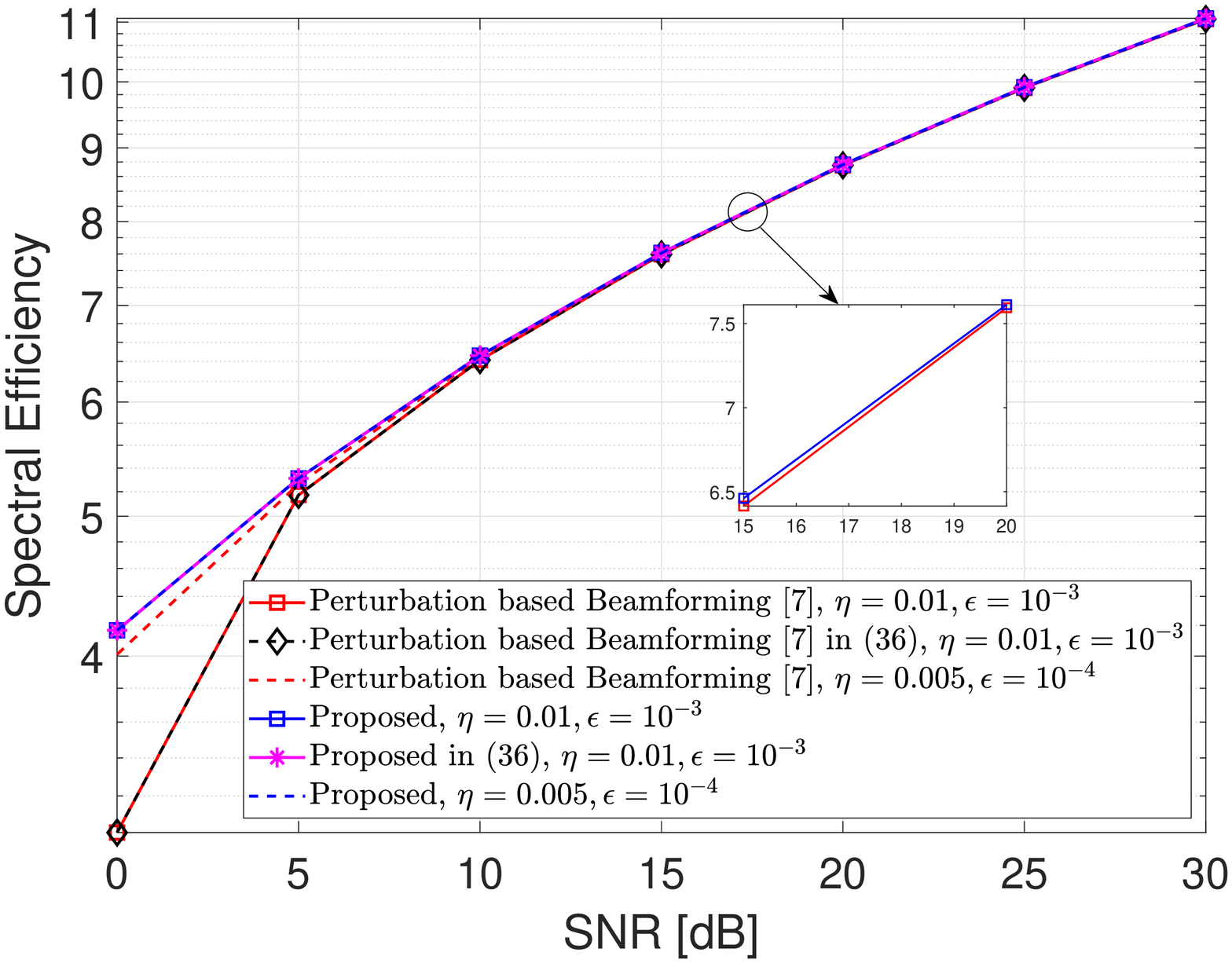}}
    \caption{Performance comparison of channel estimation  between the proposed scheme and the conventional adaptive scheme in hybrid beamformer.}
    \label{fig:hybridbf_mse}
\end{figure}
\begin{figure}[htbp]
    \centering
    \includegraphics[width=0.48\textwidth]{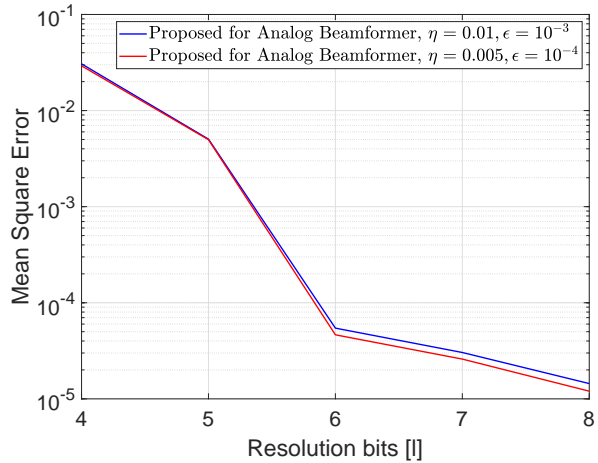}
    \caption{The mean squared error of angular parameter versus the resolution bits of phase shifter.}
    \label{fig:cdb_sz}
\end{figure}   
\begin{figure}[htbp]
    \centering
    \subcaptionbox{Mean squared error of angular parameter estimation as $\hat{u}-u$ and $\hat{v}-v$, where $u=\cos{\phi}\sin{\theta}$ and $v=\sin{\phi}\sin{\theta}$ are phase difference on the $x$-axis and $y$-axis based on $8\times8$-UPA, respectively.}{\includegraphics[width=0.48\textwidth]{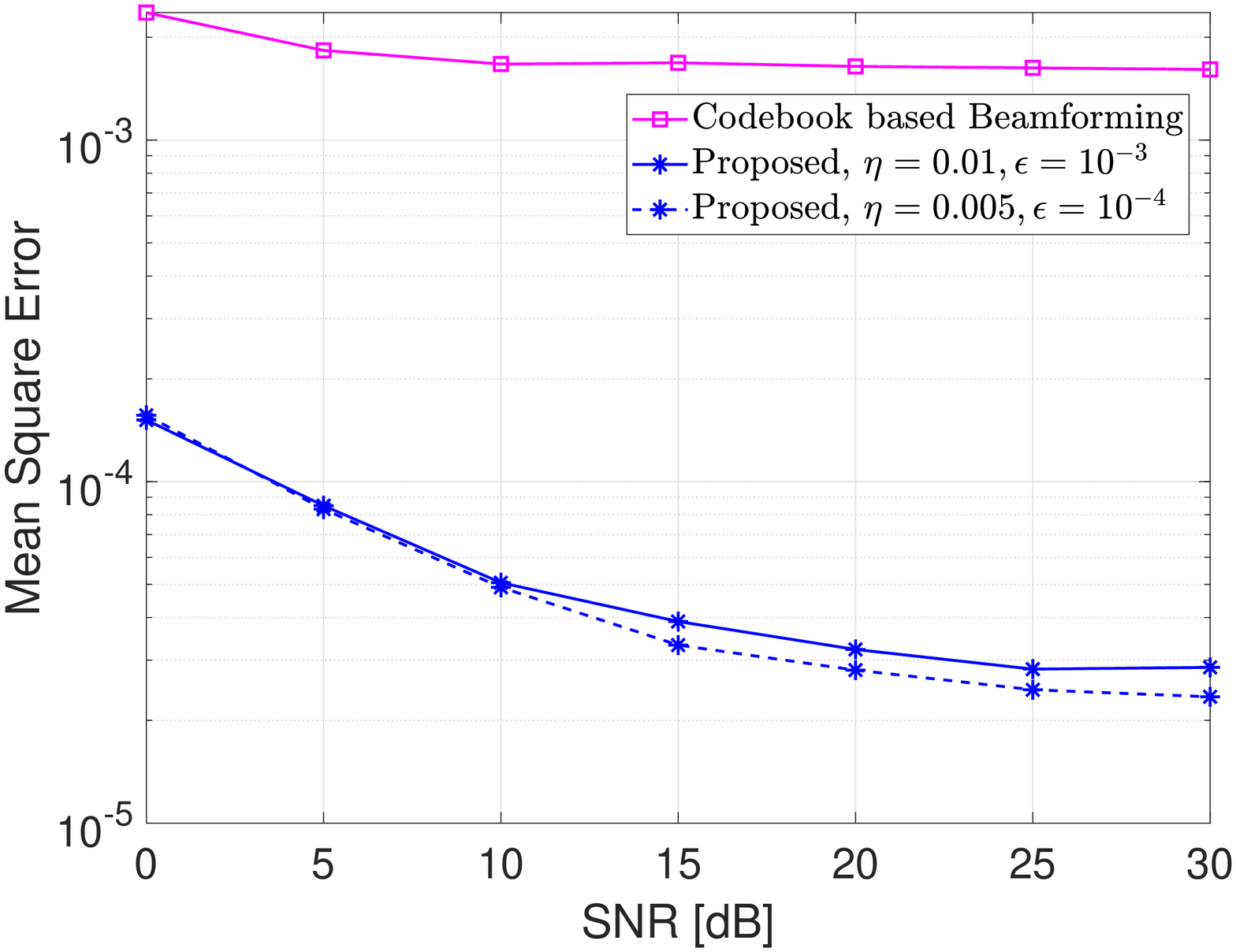}}
    \centering
    \subcaptionbox{The number of iterations for convergence.}{\includegraphics[width=0.48\textwidth]{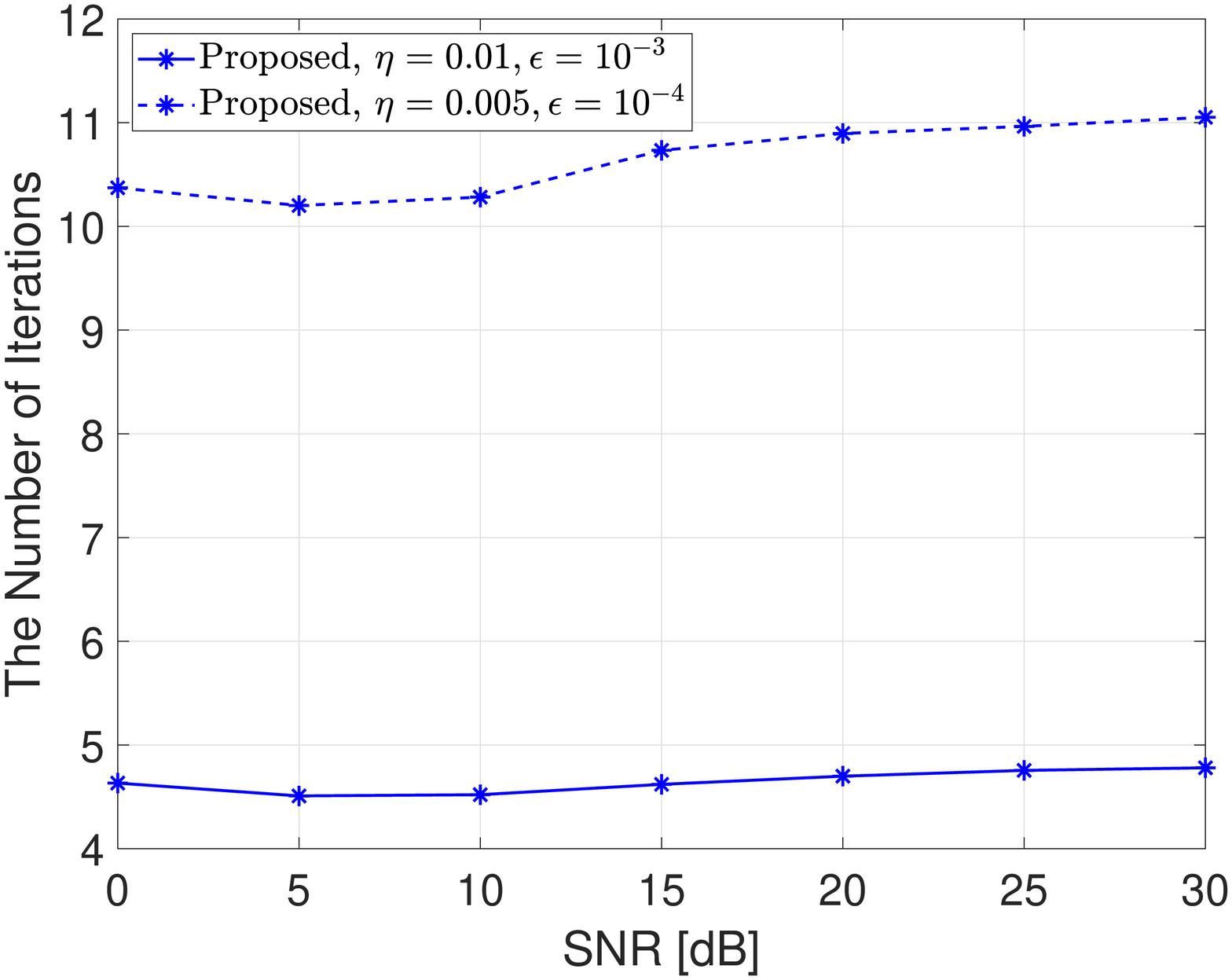}}
    \centering
    \subcaptionbox{Spectral Efficiency of directional beamforming for the estimated spatial angle as $\hat{u}$ and $\hat{v}$ based on $8\times8$-UPA.}{\includegraphics[width=0.48\textwidth]{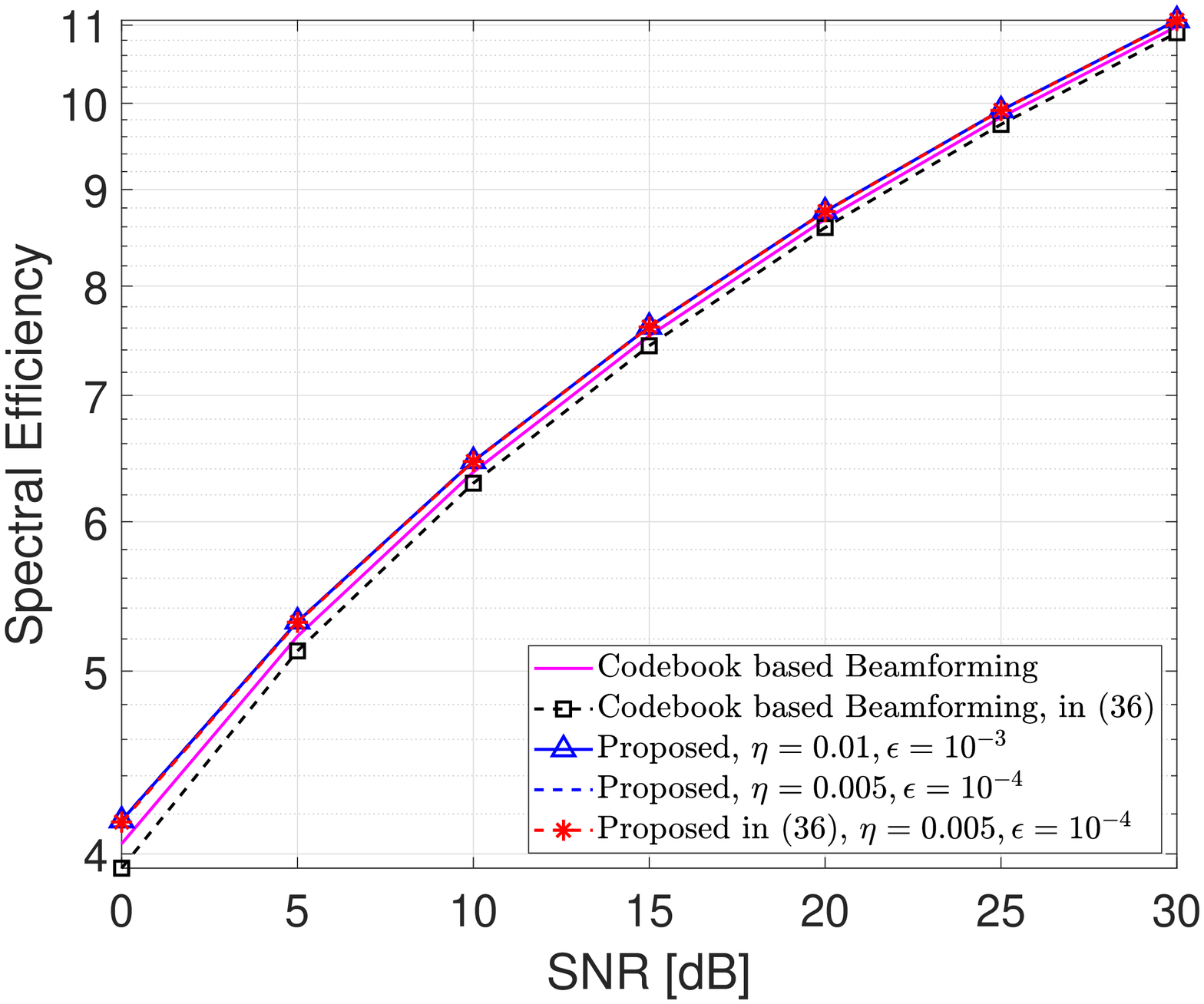}}
    \caption{Performance comparison of channel estimation  between the proposed scheme and the codebook beamforming scheme in analog beamformer.}
    \label{fig:analobf_mse}
\end{figure}
In Fig. \ref{fig:performance_tracking}, we show the trend of the normalized beamforming gain and SE over time for the proposed and other beam tracking schemes generated through Monte Carlo simulations. In Figs. \ref{fig:performance_tracking} (a) and (b), we present the performance trend assuming a high sensing accuracy case as $\sigma_G=2$ [m], $\sigma_{G,ins}=1$ [m], and $\sigma_I=0.01^\circ$. While Figs. \ref{fig:performance_tracking} (c) and (d) show the performance trend assuming a poor sensing accuracy case as $\sigma_G=5$ [m], $\sigma_{G,ins}=5$ [m], and $\sigma_I=0.05^\circ$. 
We can see from Figs. \ref{fig:performance_tracking} (a) and (b) that the GPS-only scheme is comparable to the proposed when assuming the high sensing accuracy. While it is noticeable that the integrated tracking schemes such as the proposed and the perturbation based beamforming schemes show improved performance in the poor sensing accuracy case in Figs. \ref{fig:performance_tracking} (c) and (d). Moreover, we confirm that the proposed scheme enhances the normalized beamforming gain by about $2.6\%$, and  SE by about $0.6\%$ from the GPS-only scheme. 
Also, regarding the communication-only schemes, we present that the proposed can increase the normalized beamforming gain at least by about $4.6\%$, and SE by about $1.1\%$.

\subsection{GPR based Angular Parameter Estimation} \label{ss:simul_ch}
In this subsection, we examine the effective channel estimation performance for the hybrid and analog beamformer. The GS node estimates the angular value of the effective channel $\bar{\mathbf{H}}_e=\mathbf{a}_g(u,v)$ using received pilot signals. We also set the standard deviation of the GPS control unit data as $\sigma_G=4$ [m].

\subsubsection{Hybrid Beamformer}
In Fig. \ref{fig:hybridbf_mse}, we compare our proposed scheme with the perturbation based adaptive beaforming presented in \cite{fakharzadeh2009fast,zhao2018beam,zhao2014adaptive}. Fig. \ref{fig:hybridbf_mse} shows the MSE, the number of iterations, and the SE performance as a function of SNR. 
It can be seen from Figs. \ref{fig:hybridbf_mse} (a) and (b) that the proposed scheme achieves lower MSE for angular value than perturbation based beamforming scheme with much fewer iterations. 
It should be noted that in Fig. \ref{fig:hybridbf_mse}, the proposed method with $\eta=0.005$ achieves higher MSE than that with $\eta=0.01$, despite having more iterations, which comes from the over-fitting issues. Note that as the update continues, the training data becomes large, then over-fitting issue usually occurs. Thus we need to adjust the training data size or parameter not to generate performance degradation, and we leave this issue out of scope in this paper.

We should remark that the proposed method can reduce pilot overhead requiring for angular parameter update in (\ref{eq:update}). The proposed scheme should measure the power of the received beamformed signal toward the updated angle, which requires the single measurement step. 
However, the perturbation based beamforming scheme should measure the received power twice more to calculate the gradient and need to measure the final received beamformed signal toward the updated angle \cite{fakharzadeh2009fast,zhao2018beam,zhao2014adaptive}.

In Fig. \ref{fig:hybridbf_mse} (c), we present the simulation results of SE and the calculated results in (\ref{eq:se}) from MAE. Note that we can calculate the performance metric in advance if we have information on the accuracy of the angular value estimator. 
We can see that the simulation results and the estimated SE shows a coincidence and the proposed method achieves the better SE performance than the perturbation based beamforming.

\subsubsection{Analog Beamformer}
In Fig. \ref{fig:cdb_sz}, we present the MSE performance versus various codebook size. 
As expected, we can see that the performance can be improved with the increase of resolution bits of phase shifter. However, also note that the size of the codebook, $G_s$, increases as $\{2,4,6,12,22\}$. The accuracy improves significantly as $l$ increases from $4$ to $6$, but the improvements is not noticeable when $l$ larger than $6$ bits. Moreover, in the GPR method, the calculation complexity depends on the grid size $G_s$, as $\mathcal{O}(G_s)$. Considering the time-consuming issues for the phase resolution, we decide $l$ as $6$.

In Fig. \ref{fig:analobf_mse}, we compare the the number of iteration, MSE, and SE between the proposed and codebook based analog beamforming scheme. From Figs. \ref{fig:analobf_mse} (a) and (b), it can be observed that the proposed scheme can improve the estimation performance compared to that of codebook based scheme, and the proposed method can increase the accuracy by reducing the convergence parameter $\eta$ and $\epsilon$ with the trade-off between convergence speed and the accuracy. It can be seen that the number of iterations is stable over the considered SNR region of our interest such as $0\sim30$ [dB] because of invariant training data set not by updating.
Moreover, we can observe in Fig \ref{fig:analobf_mse} (c) a slight estimation loss between the value calculated in (\ref{eq:se}) and the simulation value when SNR in low SNR region, but this is because the MAE value is over the feasible range, that is, the main lobe width.

\section{Conclusions} \label{s:conclusion}
In this paper, we proposed a flight sensor data and beamforming signal based integrated UAV tracking scheme to provide the stable link connection with the UAV node. The proposed scheme offers a compatible integrated scheme considering both the flight sensor system and the communication system. In particular, the UAV node performs the precoding using the flight sensors, and the GS node estimates the accurate position by exploiting the flight sensor measuements and beamforming signal. Moreover, we proposed a new GPR based channel estimation scheme, which can be employed in the hybrid and analog beamformers. Therefore, the proposed scheme can serve the UAV node with a persistent beam alignment, and compensate for the drawbacks of the GPS-only or the communication-only system. Simulation results showed that the proposed scheme improves the accuracy for tracking and performance metrics compared to the single schemes. Furthermore, the beamforming performance metrics of the data transmission block can be validated in advance from MAE of channel estimator.
    
\bibliography{IEEEabrv,biblist}
\bibliographystyle{IEEEtran}

\end{document}